\documentclass[twocolumn,iop,nofootinbib,reprint]{revtex4} %twocolumn,superscriptaddress,showkeys,showpacs,nofootinbib
\usepackage{graphicx}% Include figure files
\usepackage{dcolumn}
\usepackage{epsfig}
\usepackage{amsmath}

\begin{document}

\newcommand{\alumina}[0]{Al$_2$O$_3$}
\newcommand{\Kalumina}[0]{$\kappa$-Al$_2$O$_3$}
\newcommand{\Aalumina}[0]{$\alpha$-Al$_2$O$_3$}
\newcommand{\Galumina}[0]{$\gamma$-Al$_2$O$_3$}
\newcommand{\bea}[0]{\begin{eqnarray}}
\newcommand{\eea}[0]{\end{eqnarray}}
\newcommand{\nn}[0]{\nonumber}
\newcommand{\mtext}[1]{\mbox{\tiny{#1}}}
\renewcommand{\vec}[1]{{\bf #1}}
\newcommand{\op}[1]{{\bf\hat{#1}}}
\newcommand{\galpha}[0]{$\alpha$}
\newcommand{\gbeta}[0]{$\beta$}
\newcommand{\ggamma}[0]{$\gamma$}
\newcommand{\gdelta}[0]{$\delta$}
\newcommand{\gepsilon}[0]{$\eps ilon$}
\newcommand{\gphi}[0]{$\phi$}
\newcommand{\gvarphi}[0]{$\varphi$}
\newcommand{\geta}[0]{$\eta$}
\newcommand{\gtheta}[0]{$\theta$}
\newcommand{\gomega}[0]{$\omega$}
\newcommand{\gchi}[0]{$\chi$}
\newcommand{\gxi}[0]{$\xi$}
\newcommand{\gkappa}[0]{$\kappa$}
\newcommand{\del}[1]{\partial_{#1}}
\newcommand{\av}[1]{\langle #1\rangle}
\newcommand{\bra}[1]{\langle #1\right|}
\newcommand{\ket}[1]{\left| #1\right\rangle}
\newcommand{\bracket}[3]{\langle #1 | #2 | #3\rangle}
\newcommand{\sproduct}[2]{\langle #1 | #2\rangle}
\newcommand{\Bracket}[3]{\left\langle #1 \left| #2 \right| #3\right\rangle}
\newcommand{\sign}[1]{\mbox{sign}\left(#1\right)}
\newcommand{\mc}[3]{\multicolumn{#1}{#2}{#3}}
\newcommand\T{\rule{0pt}{2.6ex}}
\newcommand\B{\rule[-1.2ex]{0pt}{0pt}}

\title{
Understanding adhesion at as-deposited interfaces from
\textit{ab initio} thermodynamics of deposition growth: 
thin-film alumina on titanium carbide
}

\author{Jochen Rohrer}
\email{rohrer@chalmers.se}
\affiliation{%
BioNano Systems Laboratory,
Department of Microtechnology,
MC2,
Chalmers University of Technology,
SE-412 96 Gothenburg
}%
\author{Per Hyldgaard}%
\affiliation{%
BioNano Systems Laboratory,
Department of Microtechnology,
MC2,
Chalmers University of Technology,
SE-412 96 Gothenburg
}%

\date{\today}

\begin{abstract}
We investigate the chemical composition and adhesion 
of chemical vapour deposited thin-film alumina on TiC
using  and extending a recently proposed \textit{nonequilibrium} 
method of \textit{ab initio} thermodynamics of deposition growth (AIT-DG)
[Rohrer J and Hyldgaard P 2010 \textit{Phys.\ Rev.\ B} \textbf{82} 045415].
A previous study of this system
[Rohrer J, Ruberto C and Hyldgaard P 2010 \textit{J.\ Phys.: Condens. Matter} \textbf{22} 015004]
found that use of \textit{equilibrium} thermodynamics leads to predictions of
a non-binding TiC/alumina interface,
despite the industrial use as a wear-resistant coating. 
This discrepancy between equilibrium theory and experiment is resolved
by the AIT-DG method which predicts interfaces with strong adhesion.
The AIT-DG  method  combines  density functional theory calculations, 
rate-equation modelling  of the pressure evolution
of the deposition environment and thermochemical data.
The AIT-DG method was previously used to predict prevalent terminations of
growing or as-deposited surfaces of binary  materials.
Here we extent the method to predict surface and interface compositions
of growing or as-deposited thin films on a substrate  
and find that inclusion of the nonequilibrium deposition environment 
has important implications for the nature of buried interfaces. 
\end{abstract}


%68.35.Np Adhesion
%05.70.Np Surface and interface thermodynamics
%68.55.A- Thin film structure: Nucleation and growth
%81.15.Aa Theory and models of film growth 
\maketitle
%%%%%%%%%%%%%%%%%%%%%%%%%%%%%%%%%%%%%%%%%%%%%%%%%%%%%%%%%%%%%%%%%%%%%%%
%%%%%%%%%%%%%%%%%%%%%%%%%%%%%%%%%%%%%%%%%%%%%%%%%%%%%%%%%%%%%%%%%%%%%%%

\section{Introduction}
Interfaces and surfaces are present in practically all devices
and their detailed structure is typically crucial for the overall 
device functionality \cite{IR1, IR2, IR3,IntAdh1, SR1, SR2}.
Understanding and ultimately controlling the thin-film deposition,
the chemical composition and the adhesion at interfaces
\cite{UndAdh1,UndAdh2,UndAdh3,UndAdh4,UndAdh5,UndAdh6,UndAdh7,
UndAdh8,UndAdh9,UndAdh10,UndAdh11,UndAdh12}
is a very desirable goal of industrial and scientific research.
Characterisation of atomic structure and binding
at interfaces is a fundamental step towards this goal.

Atomistic modelling of materials 
\cite{AM1,AM2,AM3,AM4,AM5,AM6,AM7,AM8,AM9}
using methods based on \textit{ab-initio} density functional theory (DFT) 
allows for a detailed understanding of structure
at the atomic and electronic level.
Developing reliable modelling methods is of particular
value for characterisation of interfaces which
are buried inside materials
and therefore difficult to characterise 
experimentally with atomic resolution
\cite{AM2,ExpCharact1, ExpCharact2, ExpCharact3, ExpCharact4}.
Atomistic modelling has a potential to accelerate innovation,
for example, in the development of coatings and in the design of
functional surface and interface materials \cite{Innov}.
A key element of a reliable method 
is a proper treatment of thermodynamic effects 
of a surrounding environment during creation of thin films 
and interfaces.

Until recently, \textit{ab-initio} thermodynamics (AIT) methods 
were  essentially methods of surface equilibrium
(and here denoted as AIT-SE).
These methods have focused on
oxide surfaces  \cite{PhysRevB.62.4698,AIT_Scheffler} 
or metal/oxide interfaces \cite{PhysRevB.70.024103} 
assuming equilibrium between the oxide surface (interface) and O$_2$
in an O$_2$-dominated, \textit{e.g.}, ambient environment.
Oxide surfaces are in direct contact with the environment 
and the oxygen content of this environment will therefore
have a strong influence on the termination of the oxide.
However, it is by no means clear how a surrounding could 
easily influence the composition at interfaces
(which are by definition buried and insulated from the gaseous environment).
In fact, we have shown \cite{rohrer_TiC/Alumina-structure}
that adapting and applying this AIT-SE method to the TiC/alumina
interface predicts a structure and composition
that possesses essentially no binding across the interface.
This result is evidently in conflict with the actual use
of chemical vapour deposited (CVD) TiC/alumina multilayers 
as wear-resistant coating on cemented-carbide cutting-tools \cite{Halvarsson1993177, Sead}.

The present paper demonstrates that realistic descriptions of deposition environments 
are crucial for characterising  thin-film and interface compositions and, as a consequence, 
adhesion to the underlying substrate.
We show how effects  of a steady-state deposition environment (for example CVD)
can be embedded into atomistic modelling.
We employ a method of \textit{ab-initio} thermodynamics
of deposition growth (AIT-DG) \cite{rohrer_AIT-DG_TiX},
but extend it here to suit the more complex problem of
understanding thin-film formation and corresponding as-deposited interfaces.
We focus our discussion on TiC/alumina.

The fundamental strategy is to compare free energies of reaction 
associated with thin-film configurations that
differ in their detailed chemical composition.
The key elements of the method are:
(i)~analysis of free energies of reaction $G_{\text{r}}$ and
(ii)~use of rate-equation modelling for the pressure evolution of the deposition environment.
The key variables that determine  $G_{\text{r}}$ are the partial pressures 
of the various constituents of the environment.
Assuming a steady state, all partial pressures can be expressed in terms of a few rates.
No assumptions about equilibrium between some of the
species enter into this analysis.
We point out that a steady state does not necessarily
correspond to a state, sometimes described as dynamic equilibrium \cite{chemicalReactionTheory},
where the system is assumed to gain 
no free energy by the deposition  of \textit{stoichiometric} alumina.
The method also describes the evolving system 
in a certain range where dynamic equilibrium is not maintained.

The paper is organised as follows.
In Section~\ref{Sec:Background} we summarise the results of a previous structure search
and equilibrium-thermodynamic analysis for TiC/alumina interfaces \cite{rohrer_TiC/Alumina-structure}.
This section  highlights the inconsistency in adhesion properties 
of the predictions made  with the AIT-SE method.
In Section~\ref{Sec:AITDG} we give a general motivation for the use of Gibbs 
free energies of reaction to predict the prevalence of 
thin-film classes and chemical compositions at as-deposited interfaces.
Section~\ref{Sec:CVD} presents a  simple model that describes the environment
used for chemical vapour deposition of alumina on TiC.
Section~\ref{Sec:Gr} contains the details for the evaluation of Gibbs
free energies of reaction for TiC/alumina interfaces.
In  Section~\ref{Sec:Results}, we present and discuss our results.
We summarise our work and conclude in Section~\ref{Sec:Conclusions}.

\begin{figure}
\begin{tabular}{lcr}
\includegraphics[width=4.1cm]{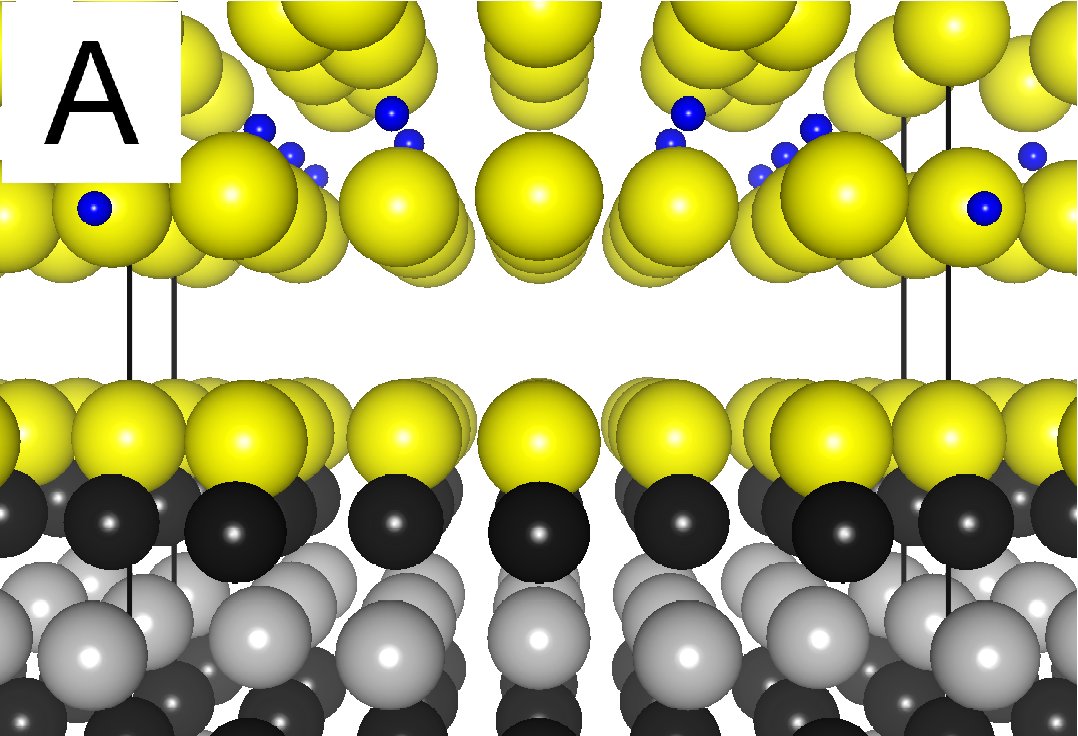}&&
\includegraphics[width=4.1cm]{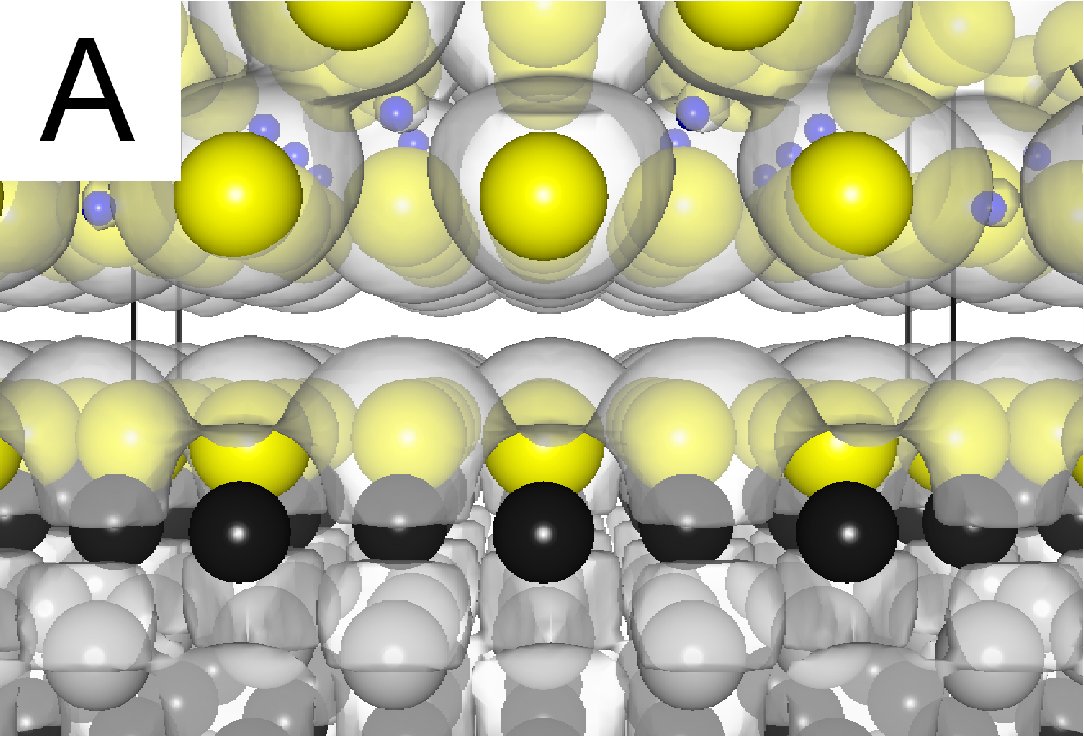}\\[0.2cm]
\includegraphics[width=4.1cm]{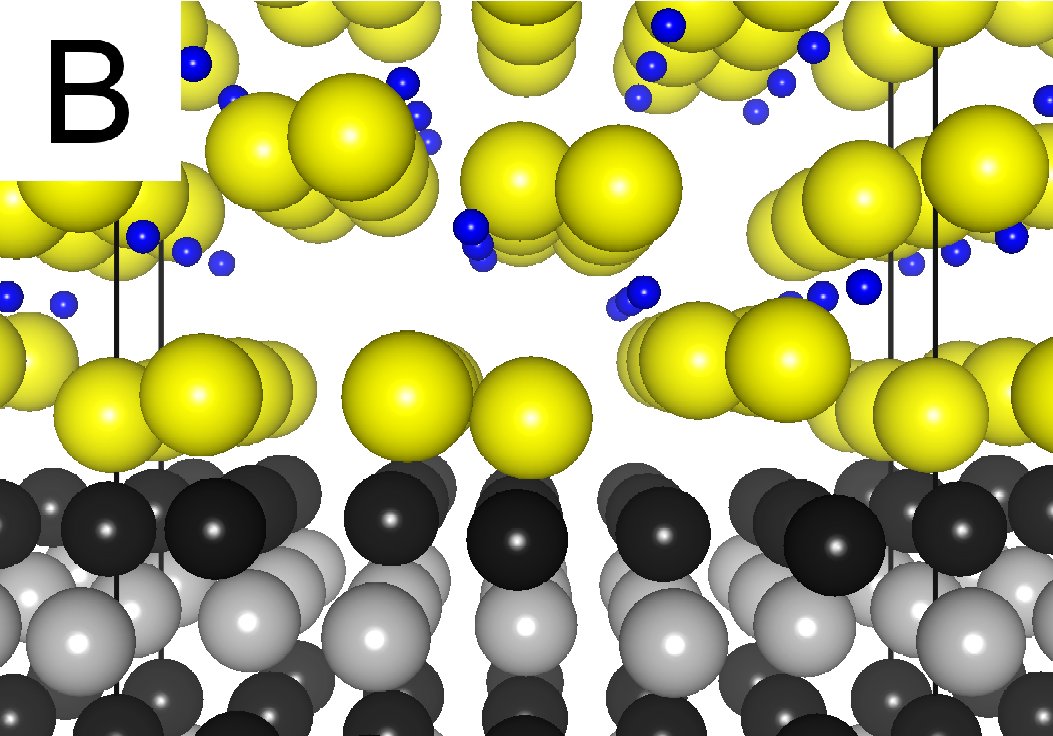}&&
\includegraphics[width=4.1cm]{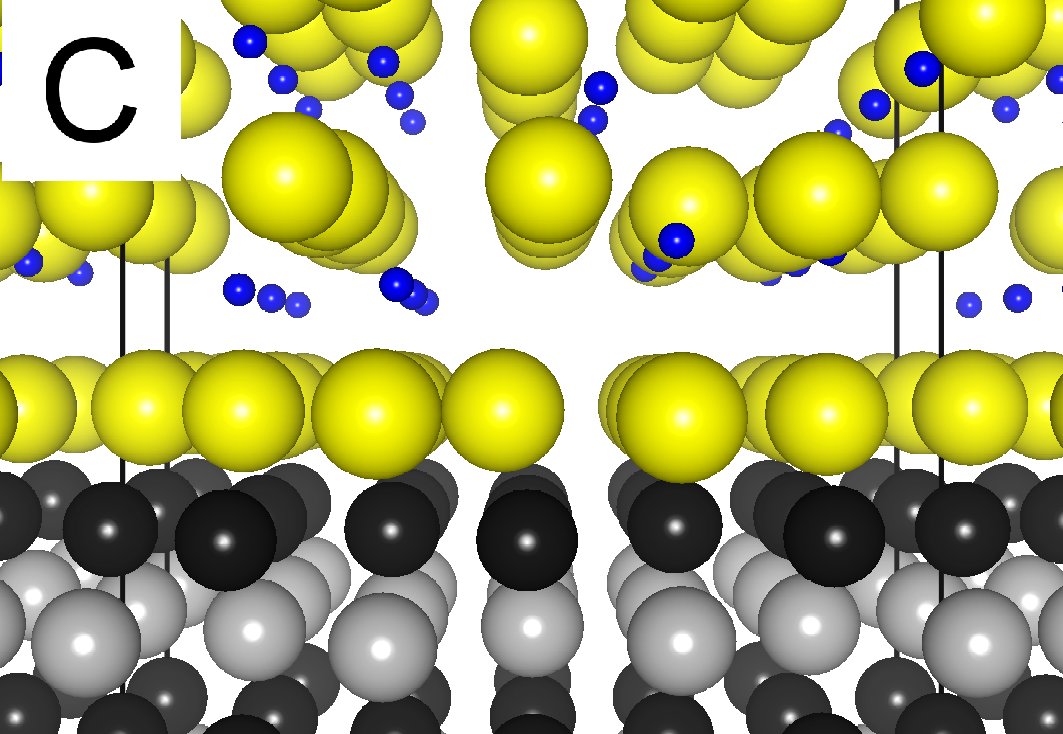}
\end{tabular}
\caption{
\label{fig:structures}
(Colour online)
Structure of interfaces between TiC and various thin-film alumina overlayers.
Colour coding: Ti = black, C = gray, Al = blue (black, small balls)
and O = yellow (light gray).
The overlayers can be sorted into classes of thin-films
with different nature of interfaces (A, B, C)
and possessing different deviations from the full \alumina\ stoichiometry
(here, Al deficiency only).
The  interface class predicted by the equilibrium AIT-SE method is shown in the top panels.
Its non-binding character is obvious from the geometric structure (left) 
and electronic density (right),
and in conflict with the wear-resistance of TiC/alumina coatings.
The set of bottom panels shows alternative
classes of thin-film alumina on TiC.
These show stronger adhesion at the interface
and have correct nature according to industrial use.
They  are found unstable according  AIT-SE 
but stable according to the present nonequilibrium account, AIT-DG.
}
\end{figure}

\section{Background \label{Sec:Background}}
In Ref.~\onlinecite{rohrer_TiC/Alumina-structure}, we have 
presented an \textit{ab initio} structure search for thin-film alumina on TiC.
We have considered thin-film alumina configurations with different thicknesses
and various chemical compositions.
The configurations can be characterised by two numbers $t$ and $\Delta_{\text{Al}}$.
Here, $t$ is the film thickness in terms of the number of O layers
and $\Delta_{\text{Al}}=N_{\text{Al}}-2/3N_{\text{O}}$
is the number of excess ($\Delta_{\text{Al}}>0$)
or deficiency ($\Delta_{\text{Al}}<0$) in the number of Al atoms.

All configurations approach the full stoichiometry in the thick-film limit.
They essentially differ in chemical composition only at the interface, surface, or both.
We therefore, in general, group the configurations into interface classes according to their
corresponding value of $\Delta_{\text{Al}}$.
Interface class A corresponds to $\Delta_{\text{Al}}=-4$,
interface class B corresponds to $\Delta_{\text{Al}}=-2$
and 
interface class C corresponds to $\Delta_{\text{Al}}=0$.
Other values of $\Delta_{\text{Al}}$ were not considered.

Figure \ref{fig:structures} presents energetically optimised 
TiC/alumina configurations for all three interface classes.
The set of top panels details of the atomic (left)
and electronic structure (right) 
at the interface class A.
Straightforward adaption and use of the \textit{equilibrium} AIT-SE method \cite{PhysRevB.62.4698,AIT_Scheffler,PhysRevB.70.024103} 
identifies interface class A as thermodynamically stable
over a wide range of temperatures and O$_2$ pressures,
see Ref.~\onlinecite{rohrer_TiC/Alumina-structure}.
However, no appreciable adhesion is found at interfaces of this type.
These interfaces separate into a TiC substrate covered with a full layer of O (TiC/O) 
and a fully stoichiometric alumina overlayer.
The electronic structure shows that the electron density essentially vanishes
between the TiC/O and the alumina,
ruling out a significant covalent binding.
Furthermore, the Al ions at the interface relax into
the first O layer above the TiC/O,
ruling out significant ionic binding.

In the present work we quantify the (lack of) adhesion by calculating the ideal work of adhesion as
$W_{\text{adh}}=(E_{\text{substrate}}+E_{\text{alumina}}-E_{\text{TiC/alumina}})/S$.
Here  $E_{\text{substrate}}$ and $E_{\text{alumina}}$
are the energies of the isolated relaxed substrate 
(TiC for interface class B and C or TiC/O for interface class A)
and the isolated relaxed alumina film 
(with in-plane lattice parameters constraint to the surface lattice of TiC).
$E_{\text{TiC/alumina}}$  is the energy of the relaxed interface
and $S$ is the area of the contact surface.
For interface class A we find a nonquantifiable (vanishing) value of $W_{\text{adh}}$,
comparable to the uncertainty in the force relaxation in the underlying 
GGA calculations (described in Ref.~\onlinecite{rohrer_TiC/Alumina-structure}).

The non-binding character at interface class A is clearly in conflict 
with the wear-resistance of TiC/alumina multilayers \cite{Halvarsson1993177}.
We attribute this conflict to the fact 
that alumina  does neither maintain equilibrium with O$_2$ during 
deposition\footnote{
At best the alumina may maintain dynamic equilibrium
with a number of gases in the surrounding.
However, as we will also discuss below,
even this assumption is too optimistic for the
case of CVD of alumina on TiC.}
nor can one expect that it reaches such equilibrium
after being removed from the deposition chamber.
This true for the present focus on thin films
and even more so for thicker overlayers.

According to the AIT-SE method,
interface class B (see bottom left  of Figure \ref{fig:structures}) 
is predicted  to be stable only  under extreme conditions,
whereas interface class C (see bottom right panel)
is predicted unstable over the whole range of allowed
values of O chemical potentials \cite{rohrer_TiC/Alumina-structure}.
However, the works of adhesion associated with these interface classes,  
$W_{\text{adh}}=7.4$ J/m$^2$ for interface class B and 
$W_{\text{adh}}=7.3$ J/m$^2$ for interface class C,
are in much better agreement with the wear-resistance of the material.
We show below that the conditions prevailing during CVD of alumina 
allow for nucleation of either one of these binding interface classes B and C.

\begin{figure}
\includegraphics[width=8cm]{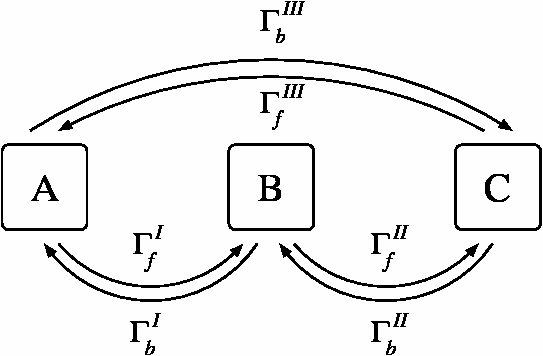}
\caption{
\label{fig:ReactionSchematics}
Schematics of elementary growth processes in an ensemble of systems
that can be organised into three different classes A, B and C.
The members of these classes could be alumina films
(of various thicknesses but) sorted by their balance of Al and O atoms
and therefore having  different surfaces and/or interfaces to the substrate.
The resulting interface classes
may exhibit a very different nature of adhesion.
Combining chemical reaction theory with a 
rate-equation description of the system allows us
to predict the probabilities for realisation of
class A, B and C during steady-state growth.
}
\end{figure}

\section{\textit{ab initio} thermodynamics of deposition growth \label{Sec:AITDG}}
In Ref.~\onlinecite{rohrer_AIT-DG_TiX} we have introduced the AIT-DG method
and demonstrated that the Gibbs free energy of reaction can be used as a predictor
for the prevalence of the chemical composition at a growing surface.
This was done for a binary material with two possible surface terminations.
Here we extend this nonequilibrium description 
to the problem of identifying the composition of growing overlayers 
which exhibit both a surface and an interface to the substrate
on which they are deposited.
We follow a similar line of argumentation as in
Ref.~\onlinecite{rohrer_AIT-DG_TiX}
and combine chemical reaction theory \cite{chemicalReactionTheory},
with a rate-equation description
of the probabilities for finding any of the possible film compositions.

Figure \ref{fig:ReactionSchematics} illustrates a collection
of systems (\textit{e.g.} an ensemble of growing thin films)
that are grouped into three thin-film and interface classes A, B and C .
The members of the classes can have different  film thicknesses but are sorted
according to their chemical composition in terms of excess or deficiency atoms of a specific species.
Members of different classes are allowed to transform into members of other classes
via the three reactions labeled as $I$, $II$, and $III$,
all being characterised by forward and backward rates
$\Gamma_f^i$ and $\Gamma_b^i$ ($i= I, II, III$).
Growth of, for example, a film of class A results by a net flow along
the reaction chain, A$\rightarrow$B$\rightarrow$C$\rightarrow$A.

Chemical reaction theory relates the forward and backward rates to 
the inverse temperature $\beta$ (in units of energy)
and  the Gibbs free energy of reaction by
$\beta \Delta G_r^i=-\ln \Gamma_f^i/\Gamma_b^i$.
At the same time, the probability for a random member of the ensemble to belong
to one of the three classes is described in terms of rate equations,
\begin{subequations}
\label{eq:RateP}
\begin{align}
\partial_t P_{\text{A}}&= -\left(\Gamma^I_f+\Gamma^{III}_b\right) P_{\text{A}}+
\Gamma^I_b P_{\text{B}}+\Gamma^{III}_f P_{\text{C}}\\
\partial_t P_{\text{B}}&=  \Gamma_f^I P_{\text{A}}-\left(\Gamma^I_b+\Gamma^{II}_f\right) P_{\text{B}}
+\Gamma^{II}_b P_{\text{C}}\\
\partial_t P_{\text{C}}&= \Gamma^{III}_b P_{\text{A}}+\Gamma^{II}_f P_{\text{B}}
-\left(\Gamma^{III}_f+\Gamma^{II}_b\right)P_{\text{C}}.
\end{align}
\end{subequations}
This follows from a straightforward generalisation of the analysis presented in 
Ref.~\onlinecite{rohrer_AIT-DG_TiX}.
The steady-state solutions for the probabilities can be expressed as ratios of 
sums of products of reaction rates, for example, 
\begin{align}
\label{eq:FullPredictor}
\frac{P_{\text{A}}}{P_{\text{B}}}=&
\frac{\Gamma_b^{I}\Gamma_b^{II}+\Gamma_b^{I}\Gamma_f^{III}+\Gamma_f^{II}\Gamma_f^{III}}
{\Gamma_b^{II}\Gamma_b^{III}+\Gamma_b^{II}\Gamma_f^{I}+\Gamma_f^{I}\Gamma_f^{III}}.
\end{align}

We use the differences in Gibbs free energies of reaction
as  an approximate predictor for the prevalence of the different 
classes of films and interfaces,
\begin{subequations}
\label{eq:Probabilities}
\begin{align}
\label{eq:p_A}
\beta \Delta G_r^I&=\ln \frac{\Gamma_b^I}{\Gamma_f^I}\approx
\ln\frac{P_{\text{A}}}{P_{\text{B}}}\\
\beta \Delta G_r^{II}&=\ln \frac{\Gamma_b^{II}}{\Gamma_f^{II}}
\approx\ln\frac{P_{\text{B}}}{P_{\text{C}}}\\
\label{eq:p_C}
\beta \Delta G_r^{III}&=\ln \frac{\Gamma_b^{III}}{\Gamma_f^{III}}
\approx\ln\frac{P_{\text{C}}}{P_{\text{A}}}.
\end{align}
\end{subequations}
The evaluations of relative probabilities (\ref{eq:Probabilities}) are exact 
in the limit where $\Delta G_r^{I}+\Delta G_r^{II}+\Delta G_r^{III}=0$
or, equivalently,  $\Gamma^I_f\Gamma^{II}_f\Gamma^{III}_f=\Gamma^I_b\Gamma^{II}_b\Gamma^{III}_b$.
This limit corresponds to dynamic equilibrium in stoichiometric growth
of thin films (irrespective to which interface class they belong).
We note that these predictors [and the full evaluation (\ref{eq:FullPredictor})]
are beyond a simple assumption of detailed balance
because the probabilities are steady-state
but not equilibrium probabilities.

With any use of the presented AIT-DG modelling, one should always check
the quality of the approximative predictors (\ref{eq:Probabilities}). The appendix
provides the theoretical framework for such a test in the case of
alumina thin-film deposition and interface formation on TiC.

\section{Modelling of CVD of alumina\label{Sec:CVD}}
\begin{figure}[t]
\includegraphics[width=8.5cm]{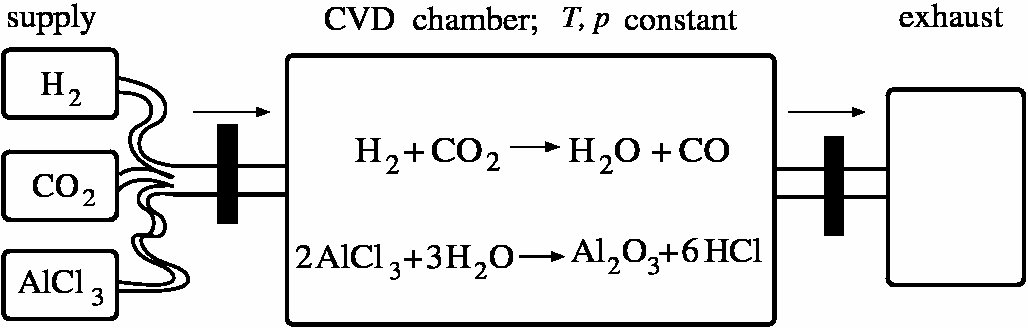}
\caption{
\label{fig:CVD}
Schematic model of chemical vapour deposition of alumina.
A H$_2$-AlCl$_3$-CO$_2$  gas mixture with relative concentrations
$c_{\text{H$_2$}}$, $c_{\text{AlCl$_3$}}$ and $c_{\text{CO$_2$}}$
is supplied at rate $R_{\text{S}}$ to a hot reaction chamber
which is kept at constant temperature.
At the same time, there is a continuous exhaust of gases
at rate $R_{\text{E}}$, also keeping the pressure constant.
Including the reactions (\ref{eq:water_production}) and (\ref{eq:CVDalumina}), 
taking place inside the chamber,
we model this system by the set of rate equations (\ref{eq:RateEquation}).
The resulting steady-state partial pressures
can be used to determine the free energies of reactions (\ref{eq:Hform})
for different TiC/alumina systems.
}
\end{figure}

\subsection{Rate-equation modelling}
Figure \ref{fig:CVD} illustrates the  CVD process utilised for alumina growth on TiC.
A H$_2$-AlCl$_3$-CO$_2$  supply gas mixture with relative concentrations $c_i$
is supplied to a hot chamber at rate $R_{\text{S}}$.
Inside the chamber water and alumina form (in parallel) 
at rates  $R_{\text{H$_2$O}}$ and  $R_{\text{Al$_2$O$_3$}}$ according to \cite{Ruppi200150}
\begin{subequations} 
\begin{align}
\label{eq:water_production}
\text{H$_2$}+\text{CO$_2$}
&\xrightarrow{R_{\text{H$_2$O}}}\text{H$_2$O}+\text{CO},\\
\label{eq:CVDalumina}
2\text{AlCl$_3$}+3\text{H$_2$O}
&\xrightarrow{R_{\text{Al$_2$O$_3$}}}
\text{\alumina}+6\text{HCl}.
\end{align}
\end{subequations}
The total pressure inside the chamber is kept at a constant value
by a continuous exhaust at rate $R_{\text{E}}$ of both reaction products
and unused supply gases.

We describe the evolution of the environment by individual partial (ideal gas) pressures 
with a  coupled set of rate equations,
\begin{align}
\partial_tp_i
\propto
c_iR_{\text{S}}-\frac{p_i}{p}R_{\text{E}}
+\nu_i^{\text{H$_2$O}}R_{\text{H$_2$O}}
+\nu_i^{\text{Al$_2$O$_3$}}R_{\text{Al$_2$O$_3$}}.
\label{eq:RateEquation}
\end{align}
Here, $p_i=p_i(t)$ is the 
momentary pressure of chemical species $i$ 
inside the reaction  chamber,
$p=p(t)=\sum_ip_i(t)$ is the momentary total pressure,
$c_i$ is the concentration of the chemical species $i$ in the supply gas,
and $\nu_i^{\text{H$_2$O}}$ and $\nu_i^{\text{Al$_2$O$_3$}}$
are the stoichiometric coefficients\footnote{
Stoichiometric coefficients are counted negative if a species is consumed
and positive if a species is produced in a reaction.}
of the chemical species $i$ in  reaction
(\ref{eq:water_production}) and (\ref{eq:CVDalumina}), respectively.

The resulting steady-state partial pressures ($\partial_t p_i(t) =0$, $p(t)=p=const.$) 
can be expressed in terms of scaled reaction rates
$r_{\text{H$_2$O}}=R_{\text{H$_2$O}}/R_{\text{S}}$
and $r_{\text{Al$_2$O$_3$}}=R_{\text{Al$_2$O$_3$}}/R_{\text{S}}$,
\begin{align}
p_i=p
\frac{c_i+r_{\text{H$_2$O}}\nu_i^{\text{H$_2$O}}
+r_{\text{Al$_2$O$_3$}}\nu_i^{\text{Al$_2$O$_3$}}}
{1+r_{\text{Al$_2$O$_3$}}}.
\label{eq:steady-state-pressures}
\end{align}
This result allows us to determine free energies of reaction
(see below) from a few experimentally controllable ($p$, $T$, $R_{\text{S}}$)
or at least measurable ($R_{\text{H$_2$O}}$, $R_{\text{Al$_2$O$_3$}}$, $R_{\text{E}}$)
quantities.

\subsection{Limits on reaction rates \label{Sec:ReactionRateLimits}}
The reaction rates possess natural bounds that cannot be exceeded in steady state.
First, for each reaction an upper bound for the reaction rate is given 
by the condition of dynamic equilibrium  in this reaction.
Reaction (\ref{eq:water_production}) can proceed from left to right only
if the chemical potentials fulfil
$\mu_{\text{H$_2$}}+\mu_{\text{CO$_2$}}\geq\mu_{\text{H$_2$O}}+\mu_{\text{CO}}$.
Similarly, reaction (\ref{eq:CVDalumina}) requires
that $2\mu_{\text{AlCl$_3$}}+3\mu_{\text{H$_2$O}}\geq \Delta G_{\text{Al$_2$O$_3$}}+6\mu_{\text{HCl}}$,
where $\Delta G_{\text{Al$_2$O$_3$}}$ is the gain in free energy per stoichiometric formula of \alumina.
Dynamic equilibrium in these reactions is reached if equality holds which corresponds
to a specific values of $R_{\text{H$_2$O}}$ and $R_{\text{Al$_2$O$_3$}}$.

In the present case another bound is found 
by comparing reactions (\ref{eq:water_production}) and (\ref{eq:CVDalumina}).
The latter reaction requires three units of H$_2$O, while the former produces only one.
Thus, in steady state, 
$R_{\text{Al$_2$O$_3$}}\leq3R_{\text{H$_2$O}}=R_{\text{Al$_2$O$_3$}}^{\text{max}}$.
However, we do not expect our model to be applicable in the the limit where
$R_{\text{Al$_2$O$_3$}}\rightarrow R_{\text{Al$_2$O$_3$}}^{\text{max}}$.
In this limit alumina deposition becomes instantaneous and on average,
the water pressure will vanish.
The ideal-gas description of the environment will therefore be inappropriate
and kinetic aspects become dominating.
We choose to consider situations where $R_{\text{Al$_2$O$_3$}}$
is sufficiently separated from  $R_{\text{Al$_2$O$_3$}}^{\text{max}}$.

\section{\textit{Ab initio} evaluation of Gibbs free energies \label{Sec:Gr}}
The Gibbs free energies of reaction are calculated as follows.
We consider a general Al$_{M}$O$_{N}$ film as the product of a 
(hypothetical) chemical reaction starting from the substrate (where $M=N=0$).
Reaction (\ref{eq:CVDalumina}) describes  stoichiometric 
solidification of alumina.
In addition, the CVD environment enables the following reaction
pathways for non-stoichiometric deposition,\footnote{
We do not include the reaction
$\text{CO$_2$}\rightarrow \text{O$_{\text{exc}}$}+\text{CO}$
since the associated free energy of reaction
is higher than that associated
with (\ref{eq:excessO}).}
\begin{subequations}
\label{eq:ExcessAtoms}
\begin{align}
\label{eq:excessAl}
\text{AlCl$_3$}+\frac{3}{2}\text{H$_2$}&\rightarrow \text{Al$_{\text{film}}$}+3\text{HCl}\\
\label{eq:excessO}
\text{H$_2$O}&\rightarrow \text{O$_{\text{film}}$}+\text{H$_2$}.
\end{align}
\end{subequations}
The label 'film' indicates the incorporation into the 
alumina film on the substrate.

We define the free energy of reaction,
$G_{\text{r}}^{M,N}$,
associated with the (hypothetical direct) deposition 
of a general Al$_{M}$O$_{N}$ film,
\begin{align}
G_{\text{r}}^{M,N}
&=G_{\text{TiC/Al$_{M}$O$_{N}$}}
-G_{\text{TiC}}\nn\\
&+M\left(3\mu_{\text{HCl}}
-\frac{3}{2}  \mu_{\text{H$_2$}}
- \mu_{\text{AlCl$_3$}}\right)\nn\\
&+N\left(\mu_{\text{H$_2$}}-\mu_{\text{H$_2$O}}\right).
\label{eq:Hform}
\end{align}
Here, $G_{\text{TiC/Al$_{M}$O$_{N}$}}$ is the free energy
of the film adsorbed on the TiC substrate,
$G_{\text{TiC}}$ is the free energy of the clean (relaxed) TiC substrate.
The set of  chemical potentials  $\mu_i$ 
describe the Gibbs free energy of the various gases that contribute to the reaction.

The Gibbs free energy definition (\ref{eq:Hform}) is unambiguous
and relevant for our nonequilibrium thermodynamic account of thin-film growth
and interface formation.
Of course, no direct reaction for the deposition
described by (\ref{eq:Hform}) exists.
However, stoichiometric combination of (\ref{eq:excessAl}) and (\ref{eq:excessO})
effectively reduces to (\ref{eq:CVDalumina}).
We treat $G_{\text{r}}^{M,N}$ as a conservative function
of the state variables $M$ and $N$
(that is independent of the details of the order of the deposition steps
that result into an Al$_M$O$_N$ film).
As a consequence, we can extract the values of $\Delta G_{\text{r}}$ 
associated with the reaction that transforms a film characterised
by $M'$ and $N'$ into a film that is characterised by $M$ and $N$
as difference between $G_{\text{r}}^{M,N}$ and $G_{\text{r}}^{M',N'}$. 
These differences  allow the direct calculation
of the relative probabilities  in the limit
of dynamic equilibrium
from our set of predictors (\ref{eq:Probabilities}).

We consider alumina films of the type
Al$_{4t-4}$O$_{6t}$ (interface class A),
Al$_{4t-2}$O$_{6t}$ (interface class B)
and Al$_{4t}$O$_{6t}$ (interface class C),
where $t$ is the thickness in terms of the number of O layers.
Two times the reaction (\ref{eq:excessAl}) thus corresponds
to reaction $I$ and $II$ in Figure \ref{fig:ReactionSchematics}
and six times the reaction (\ref{eq:excessO})
corresponds to $III$.
We note that, although $I$ and $II$ are 
both formally described by  the gas reaction (\ref{eq:excessAl}),
they are different and possess different free energies of reaction
since the substrates on which they
occur and the final products are different.

We evaluate Gibbs free energies of reaction $G_{\text{r}}^{M,N}$ as 
described in Refs.~\onlinecite{PhysRevB.62.4698,AIT_Scheffler,rohrer_AIT-DG_TiX}.
The free energies of solid phases (substrate and potential thin films) 
are replaced by their DFT total energies, $G_{\text{solid}}\approx E_{\text{solid}}$.\footnote{
In fact, we correct the total energies of the films
by subtracting the strain energy of the stoichiometric
part of the film,
$E_{\text{film}}\rightarrow E_{\text{film}}-n_{\text{Al$_2$O$_3$}}\Delta_{\text{strain}}$,
where $\Delta_{\text{strain}}$ is the 
is difference  between the strained
(due to the expansion to the TiC lattice in the interface plane)
and the unstrained bulk alumina per stoichiometric unit.}
For gases, we employ the ideal-gas approximation,
\begin{align}
\mu_i(T,p_i)=\epsilon_i+\Delta_i^0(T)+k_BT\ln(p_i/p^0).
\label{eq:mu}
\end{align}
Here $\epsilon_i$ is the DFT total energy of the gas (molecule)
and $\Delta_i^0(T)$ is the temperature dependence
of $\mu_i$ at a fixed pressure $p^0$.
We use the values  tabulated in Ref.~\onlinecite{JANAF}
for $p^0=1$~atm.
For the individual partial pressures $p_i$ we use
steady-state pressures (\ref{eq:steady-state-pressures}) 
specified by CVD process.

\section{Results and Discussions\label{Sec:Results}}
\begin{figure}
\includegraphics[width=8cm]{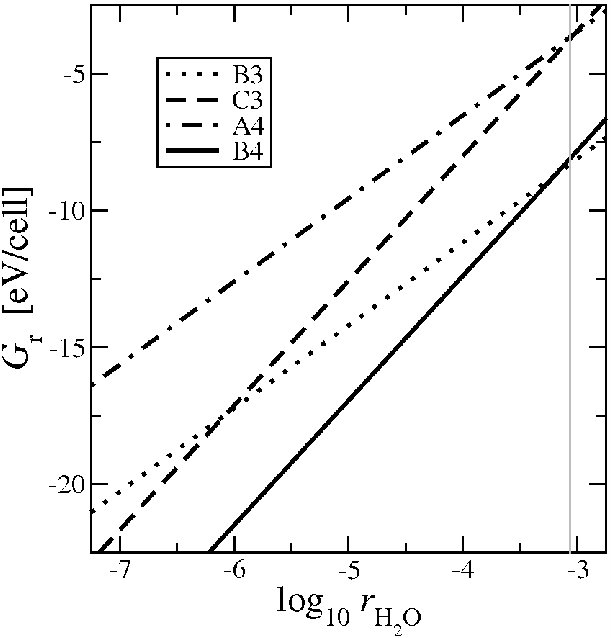}
\caption{
\label{fig:Gr}
Gibbs free energies of reaction  $G_{\text{r}}$ for formation of 
CVD thin-film alumina with different thickness and displaying different interfaces classes 
as functions of the scaled reaction rate $r_{\text{H$_2$O}}$.
The vertical line corresponds to dynamic equilibrium 
in the  water formation reaction.
Parameters of Ref.~\onlinecite{Ruppi200150} have been used for the composition of
the environment. 
A deposition temperature of $T=1000$~$^{\circ}C$, a pressure of  $p=500$~mbar
and $r_{\text{Al$_2$O$_3$}}=r_{\text{H$_2$O}}/3.1$  were assumed.
No qualitative changes arise when these parameters are varied 
in range typical for CVD of alumina \cite{Ruppi200150}.
The films are labeled according to their corresponding interface class (A, B or C)
and their thickness in terms of O layers (here the cases for 3 and 4 layers are displayed).}
\end{figure}

\subsection{Gibbs free energies of reaction and growth}
In Figure \ref{fig:Gr} we  plot the Gibbs free energies of reaction 
$G_{\text{r}}$ [see (\ref{eq:Hform})]
for various thin-film alumina configurations on TiC.
The films are labeled according to their corresponding interface class (A, B, C) and
their thickness.
We assume typical values for the the deposition temperature $T$,
the total deposition pressure $p$ and for the concentrations $c_i$ 
of the different gases in the supply gas \cite{Ruppi200150}:
$T=1000^{\circ}$C, $p=500$~mbar, $c_{\text{AlCl$_3$}}=0.04$,  $c_{\text{CO$_2$}}=0.04$, 
and H$_2$ constitutes the balance.

The vertical line corresponds to dynamic equilibrium 
in the water-producing step (\ref{eq:water_production}).
We emphasise that, in general,
dynamic equilibrium in the water-producing step
does not imply dynamic equilibrium in the alumina deposition.
In the figure we have chosen $r_{\text{Al$_2$O$_3$}}=r_{\text{H$_2$O}}/3.1$.
For this value of $r_{\text{Al$_2$O$_3$}}$ the dynamic equilibria roughly
coincide within our our approximation for the Gibbs free energy variation \cite{rohrer_TiC/Alumina-structure}.
This follows from the observation that  the values of $G_{\text{r}}$ for the B3 and B4 films 
(which differ in thickness by  one full layer or two stoichiometric units)
are approximately equal on the vertical line.
We note, however, that there is an uncertainty in this value 
of  $r_{\text{Al$_2$O$_3$}}$ for coinciding dynamic equilibria.
The reason is the uncertainty in the calculation of chemical potentials
(in particular total energies of the molecules,
see discussion in Ref.~\onlinecite{rohrer_AIT-DG_TiX})
and the uncertainty in the total energies of the films themselves.
For the latter to be accurate we would have to make sure
that their atomic structures are fully optimised.
The geometries identified in Ref.~\onlinecite{rohrer_TiC/Alumina-structure}
and used here are candidates for the optimised structures
but not guaranteed to \textit{the} optimised structures.

Figure \ref{fig:Gr} shows that,
within the possible range of $r_{\text{H$_2$O}}$,
all films have a negative value of $G_{\text{r}}$.
As a consequence all films are stable with respect to 
the substrate.
This remains true also if we decrease the value of $r_{\text{Al$_2$O$_3$}}$.

The figure also demonstrates stoichiometric growth of alumina films.
Stoichiometric growth corresponds to 
$\Delta G_r^I+\Delta G_r^{II}+\Delta G_r^{III}\leq0$.
This condition is fulfilled as can be seen from the fact that
the B4 film has lower free energy of reaction than the B3 film
and that the former consists of two stoichiometric units of alumina
more than the latter.

We emphasise that the Gibbs free energy variations
shown in figure \ref{fig:Gr}
directly reflect the closely related nature 
of process $I$ and $II$.
Process $I$ and $II$ differ only in the solid reactants and solid products,
the gaseous reactants and products are the same.
In chemical reaction theory \cite{chemicalReactionTheory}
we can express the Gibbs free energy of reaction as
$\beta \Delta G_{\text{r}}=\ln \Gamma_f/\Gamma_b=-\ln K_{\text{eq}}+\ln Q$.
Here $K_{\text{eq}}=k_{\text{f}}/k_{\text{b}}$ is the
equilibrium constant of the reaction,
$k_{\text{f}}$ ($k_{\text{b}}$) is the forward (backward) reaction rate constant
and $Q$ is the reaction quotient (ratio of concentrations of gaseous products and  of reactants).
Since the gaseous reactants and products in process $I$ and $II$ are identical,
the reaction quotients for reaction $I$ and $II$ are identical.
Therefore, 
$\beta (\Delta G_{\text{r}}^{I}-\Delta G_{\text{r}}^{II})=\ln (K_{\text{eq}}^{II}/K_{\text{eq}}^I)$
must be constant.
In figure \ref{fig:Gr} we have
\begin{align}
&\beta (\Delta G_{\text{r}}^{I}-\Delta G_{\text{r}}^{II})\nn\\
&=[G_r(B4)-G_r(A4)]-[G_r(C3)-G_r(B3)]\nn\\
&=[G_r(B4)-G_r(C3)]-[G_r(A4)-G_r(B3)].
\label{eq:Parallel}
\end{align}
We notice that the curves corresponding to B4 and C3 films are 
approximately parallel.
The same applies for the curves corresponding to A4 and B3 films.
Thus, both differences after the second equal sign in  (\ref{eq:Parallel})
are constant and the AIT-DG results can also be used
to compute $\ln (K_{\text{eq}}^{II}/K_{\text{eq}}^I)$.

\subsection{Thermodynamic analysis}
\begin{figure}
\includegraphics[width=8cm]{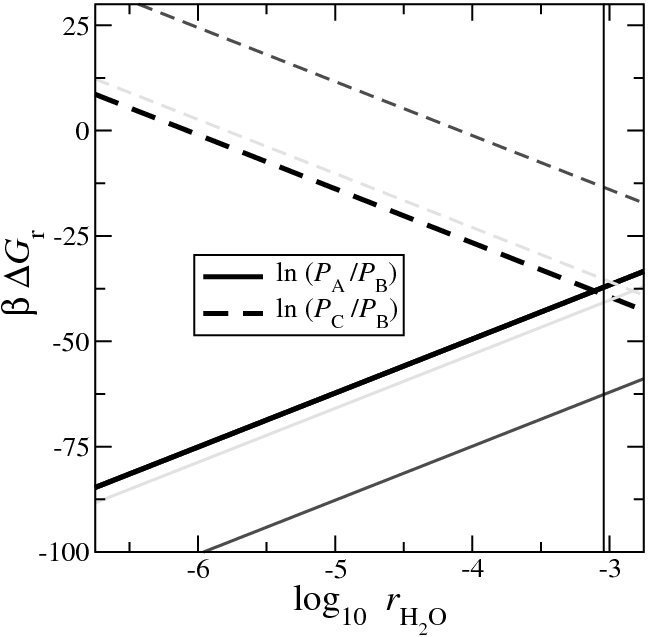}
\caption{
\label{fig:P}
Predictors for prevalence of interface class A (solid lines)
and interface class C (dashed lines)
as a functions of the scaled reaction rate $r_{\text{H$_2$O}}$
for water formation.
The predictors are the approximate logarithm of the probability relative to the probability of
prevalence of interface class B. 
The vertical line limits $r_{\text{H$_2$O}}$ to the right 
and corresponds to dynamic equilibrium in the water formation, see (\ref{eq:water_production}).
Parameters of Ref.~\onlinecite{Ruppi200150} have been used for the composition of
the environment and a deposition temperature of
$T=1000$~$^{\circ}C$ was assumed.
Black thick lines correspond to a deposition pressure of $p=500$~mbar
and $r_{\text{Al$_2$O$_3$}}=r_{\text{H$_2$O}}/3.1$,
thin light lines to $p=10$~mbar
and $r_{\text{Al$_2$O$_3$}}=r_{\text{H$_2$O}}/3.1$,
and thin dark lines to $p=500$~mbar
and $r_{\text{Al$_2$O$_3$}}=r_{\text{H$_2$O}}/300$.
Close to dynamic equilibrium limit in reaction (\ref{eq:water_production}), 
we predict the highest prevalence for interface class B.
As  $r_{\text{H$_2$O}}$ decreases, the likelihood for the occurrence
of interface class C increases.
This likelihood strongly increases further with decreasing  $r_{\text{Al$_2$O$_3$}}$
and also as the total deposition pressure increases. 
The nonbinding interface class A is highly unlikely to be created in the CVD process.
}
\end{figure}

Figure \ref{fig:P} reports the calculated predictor  (\ref{eq:Probabilities})
for the prevalence of interface class A and C relative to interface class B
as functions of the scaled reaction rate for water formation [see (\ref{eq:water_production})].
This predictor corresponds to the approximate
relative steady-state probabilities
(being exact in the limit of dynamic equilibrium
in the alumina deposition).
We have tested the quality of our predictor  
over a wide range of possible choices for 
unknown rate constants;
we refer to the appendix for a more detailed presentation.

The vertical line corresponds to dynamic equilibrium 
in the water-producing step (\ref{eq:water_production}).
We assume the same values as before for the temperature and the concentrations
of the different species in the supply gas \cite{Ruppi200150}.
For the total deposition pressure and the reaction rate of alumina deposition 
we consider the following pairs of parameters, 
$p=500$~mbar and $r_{\text{Al$_2$O$_3$}}=r_{\text{H$_2$O}}/3.1$,
$p=10$~mbar and $r_{\text{Al$_2$O$_3$}}=r_{\text{H$_2$O}}/3.1$ 
and $p=500$~mbar and $r_{\text{Al$_2$O$_3$}}=r_{\text{H$_2$O}}/300$.
The location of the vertical line is independent of these parameters
(but depends on temperature).

We find that, in or close to dynamic equilibrium in the water producing step,
the (approximate) probabilities for the prevalence of interface class A and C are much smaller
than the probability for the prevalence of interface class B,
$P_{\text{X}}<<P_{\text{B}}$ (X = A or C).
As  $r_{\text{H$_2$O}}$ decreases $P_{\text{C}}$ increases and 
becomes larger than $P_{\text{B}}$ at some point.
The value of $P_{\text{A}}$, on the other hand, decreases
as  $r_{\text{H$_2$O}}$ decreases.
Thus, our results show that the nonbinding interface class A is not realised
in CVD of alumina as described in Section~\ref{Sec:CVD}.
Instead, if reaction (\ref{eq:water_production}) takes place sufficiently close to
dynamic equilibrium,
interface class B is predicted.

We note that it is by no means obvious that dynamic equilibrium is always maintained.
If that was the case, $r_{\text{Al$_2$O$_3$}}$ would always assume its maximum value.
Then, however, it would be possible to scale the absolute deposition rate 
$R_{\text{Al$_2$O$_3$}}=r_{\text{Al$_2$O$_3$}}\cdot R_{\text{S}}$
to infinity simply by increasing the supply rate.
It is against common sense to expect the deposition 
to remain in dynamic equilibrium as we increase the supply flux.

\begin{figure}
\includegraphics[width=8.2cm]{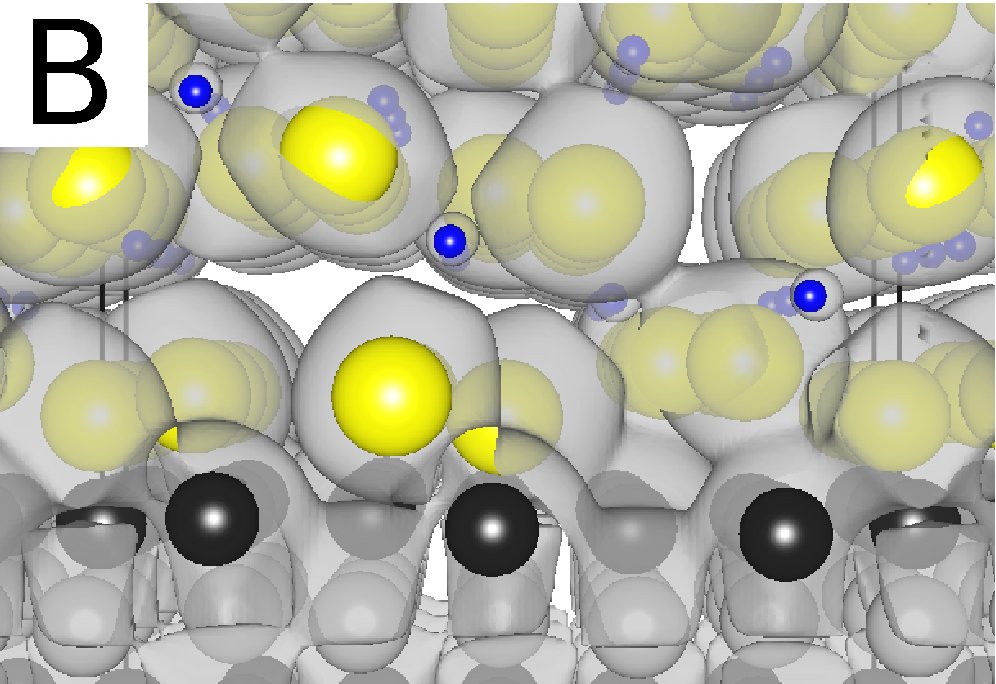}
\caption{
\label{fig:Bonding}
(Colour online)
Electron density at the interface B which shows the highest probability
to be created in CVD of alumina on TiC.
Same colour coding as in Fig.~\ref{fig:structures}.
}
\end{figure}

\subsection{Characterisation of interface adhesion}
Figure \ref{fig:Bonding} details the electronic density  at the interface class B.
We have chosen the same isosurface level as in the top right panel of  figure \ref{fig:structures}.
The Ti-O bonds show both ionic and covalent character.
The density also reveals the partial covalent character 
of the otherwise highly ionic alumina itself.
This covalent character can be seen by a small amount of electrons 
spilling over from neighbouring O layers.

A Bader analysis including core charges \cite{BaderAnalysis,BaderOyvind1,BaderOyvind2} 
shows that the ionicity of the Ti atoms  is
lower at the interface B ($q_{\text{Ti}}=1.71~e$)
than at the interface A ($q_{\text{Ti}}=1.88~e$, top panels of  figure \ref{fig:structures}).
At the same time, the ionicity of the O atoms
is higher ($q_{\text{O}}=-1.46~e$) at interface B
than at interface A ($q=-1.13~e$).
The Ti-O bond strength is lower at the  interface B
than at the interface A.
This is reflected by an increased  Ti-O layer separation
at interface B ($l_{\text{Ti-O}}=1.48$) 
with respect to  that at interface A ($l_{\text{Ti-O}}=0.88$).
However, the increased ionicity of the first O layer at interface B
and the presence of Al ions between the first and the second O layers
above the interface plane give rise to a strong overall 
interface adhesion.
This adhesion is absent at interface A.
Thus, the strong binding at the wear-resistant interface B 
that is most likely to be created in CVD of alumina 
arises from an interplay between a softening of the Ti-O bonding,
an increased ionicity in the first O layer
and a strong ionic binding in the alumina.
The latter is supplemented by weaker covalent bonds
between O layers in the alumina coating.

\section{Summary and Conclusion \label{Sec:Conclusions}}
We have extended a recently proposed \textit{nonequilibrium}
method for \textit{ab initio} thermodynamics of deposition growth 
from application to surfaces terminations  \cite{rohrer_AIT-DG_TiX} 
to prediction of thin-film and interface formation.
Our results demonstrate that a careful treatment of deposition conditions
in thin-film and interface modelling is crucial 
for understanding adhesion properties for  CVD alumina on TiC.
Assuming equilibrium between the oxide and O$_2$ (as in \cite{PhysRevB.62.4698,AIT_Scheffler,PhysRevB.70.024103})
results in prediction of a configuration that shows no binding across the interface 
(interface class A) \cite{rohrer_TiC/Alumina-structure},
see top panel in Fig.~\ref{fig:structures}.
This is in conflict with the wear-resistant nature 
and industrial use  of the material \cite{Halvarsson1993177}.
In agreement with the wear-resistance of TiC/alumina coatings,
we predict the  deposition of strongly binding
interface of type B (in or close dynamic equilibrium in the water forming step)
or interface of type C (away from dynamic equilibrium).

We expect that a similar analysis will be necessary 
also for other buried interfaces that form during a deposition
process in an environment that strongly differs 
from ambient conditions.
The thermodynamic method used here only makes reference to the molecular species that
are present (and directly relevant) during deposition.
In contrast to  equilibrium-thermodynamics approaches 
\cite{PhysRevB.62.4698,AIT_Scheffler, PhysRevB.70.024103},
this method is therefore not limited to oxides 
(although we here illustrate the method for a particular oxide)
or materials that contain a constituent X
for which a dimer X$_2$ could serve as reference.
Furthermore, the method allows for a search of conditions 
(supply gas compositions, deposition temperatures, deposition temperatures)
that favour deposition of a pre-specified interface composition
with desirable properties.
Our results suggest that the nonequilibrium 
\textit{ab initio} thermodynamics method (Ref.~\cite{rohrer_AIT-DG_TiX} 
and present extension)
can be useful  in guiding experimental optimisation of present-day materials
and design of novel such.

\section*{Acknowledgement}
We thank G. D. Mahan and C. Ruberto for valuable discussions.
Support by the Swedish National Graduate School in Materials Science (NFSM),
the Swedish Foundation for Strategic Research (SSF) through ATOMICS,
the Swedish Research Council (VR)
and the Swedish National Infrastructure for Computing (SNIC)
is gratefully acknowledged.

\begin{appendix}

\section{Determination of film prevalence}
The presentation of our results for the prevalence of the different alumina films rests on the analysis
of the predictors (\ref{eq:Probabilities}).
These predictors are strictly valid only under the assumption of dynamic equilibrium
(in the alumina deposition).
In this appendix we give a prescription of how the quality of these predictors 
can be tested with respect to the exact steady-state probability
for the prevalence of different film classes.

We consider the exact steady-state ratio between the probability for A and B,
see equation (\ref{eq:FullPredictor}).
We introduce $x_i=\Gamma_b^{i}/\Gamma_f^{i}=\exp(\beta G_r^i)$ 
and $\alpha=x_1^{-1}x_2^{-1}x_3^{-1}$.
The functions $\alpha$ and $x_i$ depend on $r_{\text{H$_2$O}}$
but are otherwise completely  determined by our \textit{ab initio} calculations 
(and the thermochemical data to determine the chemical potentials).
In particular, the parameter $\alpha$ measures the departure from dynamic equilibrium;
$\alpha=1$ corresponds to dynamic equilibrium,
$\alpha>1$ corresponds to growth beyond dynamic equilibrium
and $\alpha<1$ corresponds to evaporation.
We furthermore introduce  $F_1=\Gamma_f^{I}\Gamma_f^{II}$ and
$F_2=\Gamma_f^{III}\Gamma_f^{I}$.
These functions also depend on $r_{\text{H$_2$O}}$
but require some additional parametrisation, see below.

Independent of the parametrisations, we can express the ratio of probabilities
(\ref{eq:FullPredictor}) as
\begin{align}
\label{eq:pA}
\frac{P_{\text{A}}}{P_{\text{B}}}=&
\frac{
x_1x_2+x_1F_1F_2+F_2
}
{
\alpha^{-1}x_1^{-1}F_2+x_2+F_1F_2
}.
\end{align}

For the ratio between the probability for C and B we find
\begin{align}
\frac{P_C}{P_B}&=
\frac{
\Gamma_b^I\Gamma_b^{III}+\Gamma_f^{II}\Gamma_b^{III}+\Gamma_f^{I}\Gamma_f^{II}
}
{
\Gamma_b^{II}\Gamma_b^{III}+\Gamma_f^{I}\Gamma_b^{II}+\Gamma_f^{I}\Gamma_f^{III}
}\nn\\
&=
\frac{
F_1F_2+F_2+\alpha x_1x_2
}
{
x_2F_2
+
\alpha x_1x_2^{-1}
+
\alpha x_1x_2F_1F_2
}.
\label{eq:pC}
\end{align}

The quality of the approximate predictors (\ref{eq:Probabilities}) 
can be tested by evaluating (\ref{eq:pA}) and (\ref{eq:pC}).
The evaluation of  (\ref{eq:pA}) and (\ref{eq:pC})
requires the specification of the functions $F_1$ and $F_2$.
Here, we assume the following parametrisations,
\begin{align}
F_1&=\frac{\Gamma_f^{I}}{\Gamma_f^{II}}=
\frac{k_f^{I}}{k_f^{II}}\frac{\Pi_i [X_i^{I}]^{\nu_i^I}}{\Pi_j[X_j^{II}]^{\nu_j^{II}}}\\
F_2&=\frac{\Gamma_f^{III}}{\Gamma_f^{I}}
=\frac{k_f^{III}}{k_f^{I}}\frac{\Pi_k [X_k^{III}]^{\nu_k^{III}}}{\Pi_i[X_i^{I}]^{\nu_i^I}},
\label{eq:F2}
\end{align}
where the $\nu_i$ can but do not need to be identical 
with the stoichiometric coefficients.
We note that this parametrisation
in only a crude approximation and that growth can,
in general, be more complicated.

For alumina growth, the reactions $I$ and $II$ are formally identical.
The only difference is that they take place on different substrates
and generate different solid products.
Therefore, $F_1$ reduces to the ratio of the rate constants, 
which is constant.
Reaction $I$ and $III$ are different and $F_2$ therefore depends on $r_{H_2O}$.
Thus we have
\begin{align}
F_1&=\frac{k_f^{I}}{k_f^{II}}\\
F_2&=\frac{k_f^{III}}{k_f^{I}}\frac{[H_2O]^m[H_2]^{-n}}{[AlCl_2]^r[H_2]^s[HCl]^{-t}},
\end{align}
where $m$, $n$, $r$, $s$, $t$ are all positive and of the order of 1.

Since the rate constants $k_f^{I}$, $k_f^{II}$, and $k_f^{III}$ are unknown,
we have compared (\ref{eq:p_A}) with (\ref{eq:pA}) and (\ref{eq:p_C}) with (\ref{eq:pC}) for 
several choices of $m$, $n$, $r$, $s$, $t$ and a broad range
of ratios of rate constants.
We find that (\ref{eq:pC}) is essentially described by (\ref{eq:p_C}).
For  (\ref{eq:pA}) the approximation (\ref{eq:p_A}) can deviate as $r_{\text{H$_2$O}}$ decreases
but the qualitative result ($P_{\text{A}}<<P_{\text{B}}$) 
is not affected.
Moreover, the ratio $P_{\text{A}}/(P_{\text{B}}+P_{\text{C}})$
is always strongly suppressed.

\end{appendix}

\end{document}